\documentclass[conference]{IEEEtran}
\usepackage[top=0.75in, bottom=1.05in, left=0.625in, right=0.625in]{geometry}
\IEEEoverridecommandlockouts

\usepackage{cite}
\usepackage{amsmath,amssymb,amsfonts}
\usepackage{algorithmic}
\usepackage{graphicx}
\usepackage{textcomp}
\usepackage{xcolor}
\usepackage{url}

\def\BibTeX{{\rm B\kern-.05em{\sc i\kern-.025em b}\kern-.08em
    T\kern-.1667em\lower.7ex\hbox{E}\kern-.125emX}}
    
\usepackage{xcolor,soul,framed} 

\begin{document}
\bstctlcite{IEEEexample:BSTcontrol}
    \title{Coverage-aware and Reinforcement Learning Using Multi-agent Approach for HD Map QoS in a Realistic Environment}

\author{Jeffrey Redondo\IEEEauthorrefmark{1} (Student Member, IEEE), Zhenhui Yuan\IEEEauthorrefmark{2} (Member, IEEE),  
\\Nauman Aslam\IEEEauthorrefmark{1} (Senior Member, IEEE), Juan Zhang\IEEEauthorrefmark{1} (Member, IEEE)\\
\IEEEauthorblockA{\IEEEauthorrefmark{1} \texttt{@northumbria.ac.uk}, University of Northumbria, Newcastle, UK}
\IEEEauthorblockA{\IEEEauthorrefmark{2} \texttt{@warwick.ac.uk}, University of Warwick, Coventry, UK}}

\maketitle

\begin{abstract}
One effective way to optimize the offloading process is by minimizing the transmission time. This is particularly true in a Vehicular Adhoc Network (VANET) where vehicles frequently download and upload High-definition (HD) map data which requires constant updates. This implies that latency and throughput requirements must be guaranteed by the wireless system. To achieve this, adjustable contention windows (CW) allocation strategies in the standard IEEE802.11p have been explored by numerous researchers. Nevertheless, their implementations demand alterations to the existing standard which is not always desirable. To address this issue, we proposed a Q-Learning algorithm that operates at the application layer. Moreover, it could be deployed in any wireless network thereby mitigating the compatibility issues. The solution has demonstrated a better network performance with relatively fewer optimization requirements as compared to the Deep Q Network (DQN) and Actor-Critic algorithms. The same is observed while evaluating the model in a multi-agent setup showing higher performance compared to the single-agent setup.
\end{abstract}


\begin{figure}[!b]
    \centering
    \includegraphics[width=\columnwidth]{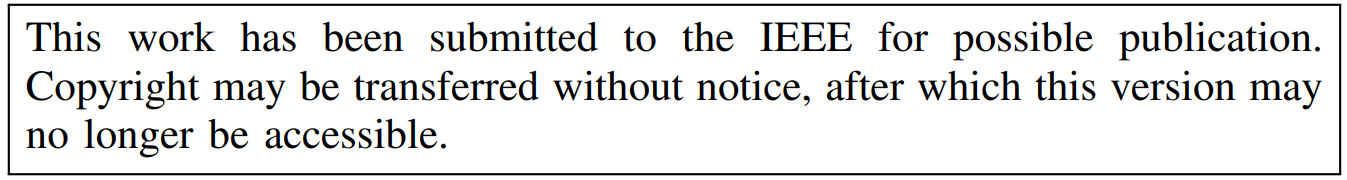}
\end{figure}

\IEEEpeerreviewmaketitle

\section{Introduction} \label{intro}
By 2025, it is expected to have autonomous vehicles (AVs) on the road \cite{gov_uk} in many developed countries around the world. However, to accomplish this, AVs must achieve the highest level of automation, Level-5 \cite{sae_level5}. To this end, the new HD map application has been developed with the potential to provide the AV with a significant amount of road information with accuracy in centimeters to guarantee a safe driving experience. The construction of the HD map is possible due to the post-processing of the raw data generated by the AV's sensors, such as cameras or LiDARs. Furthermore, the HD map application has emerged to enable the next level of automation \cite{nvidea_hdmap}. Nevertheless, Generating an HD map demands a high-powered computing capacity that can overwhelm the CPU \cite{chameleon}, and it may be inefficient for each AV to create its map. Therefore, it is suitable that the raw data must be offloaded to the edge/cloud servers. For instance in \cite{hdmap_processing_time}, results demonstrated an improvement of 66\% (from 23s to 7.8s for a 20m HD Map) for the generation of the HD Map.
Nonetheless, for a Vehicular Ad-hoc Network (VANET) that utilizes the standard IEEE802.11p, there is a problem with the fixed CW size which increases the packet collision as studied in \cite{ieee80211_cw2}. Consequently, in a high-density and high-mobility network such as VANET, there might be an increase in latency which translates into a higher processing time for the generation of HD maps. There have been efforts to improve the latency, the throughput, and the packet delivery ratio, through developing a new access category \cite{low_latency_new_ac}, utilizing machine learning algorithm such as Q-learning \cite{q_learning_fairness}, or game theory solution \cite{adaptive_edca}.

Nonetheless, those solutions require adapting or modifying the current standard which is not suitable for a worldwide network with a defined standard. Furthermore, some of these solutions are limited to specific types of services or only HD maps, which may not align with real-world scenarios where AVs may transmit various types of data, including Best-effort (BE),  video (VI), and Voice (VO). To address this, we have developed a multi-service and multi-agent system in a realistic scenario in a city in the United Kingdom. The solution operates in the application layer to mitigate the compatibility issue between wireless technologies. Additionally, we have incorporated the sojourn time which represents the time a vehicle spends in the coverage area.

\subsection{Contributions} 
\begin{itemize}
\item A comprehensive model evaluation was conducted with a more realistic traffic flow scenario derived from a real-world environment. The multi-agent solution demonstrated the robustness of the model, which is re-trained to showcase the adaptability of the RL algorithm to new scenarios. This provides valuable insights into the practical significance of real-world scenarios.
\end{itemize}

The structure of the paper is as follows: Section \ref{related_work} includes a thorough literature review on the CW, and Enhanced Distributed Channel Access (EDCA). Section \ref{problem_statement} describes the problem statement, while Section \ref{design} contains the machine learning algorithm. Section \ref{simulation} provides the assumption considered for the simulation and tools used. Section \ref{results} refers to the experiment results, and discussion. The conclusion section is Section \ref{conclusion}.

\section{Related Work} \label{related_work}
In this section, it is described the related work performed with the aim of improving network performance in terms of latency and throughput in IEEE802.11 wireless networks.

\subsection{Contention Windows \& EDCA}
Research efforts focused on improving IEEE802.11 by finding the optimal CW or modifying the Enhanced Distributed Channel Access (EDCA) access categories (ACs). One of the approaches is related to the management of the AC queues. For instance, Shinzaki et al. \cite{q_learning_edca_policy_RL} has improved the latency by mapping packets between low and high-priority ACs queues which results in an enhancement of $13.8\%$.

Moving forward, researchers have opted for modifying the current EDCA which provides new priority for specific services. For instance, a new AC for low-latency applications has been developed in \cite{low_latency_new_ac}, and others implemented an AC for each different type of video resolution \cite{avaq_edca_new_ac}. Nonetheless, this is not sufficient. As it is known each AC has a fixed CW parameter which produces higher packet collision in a dense environment as studied in \cite{ieee80211_cw2}. More specifically, in a Vehicular network as stated in \cite{performance_analysis_HDMAP} authors highlighted the negative impact the fixed CW has in the network in regards to the number of AVs and their velocity. In order to mitigate this problem, researchers have proposed new strategies for finding the optimal value of CW using machine learning paradigms. Those efforts include the use of Q-learning \cite{q_learning_fairness}, policy gradient \cite{q_learning_edca_policy_RL}, and deep RL \cite{adaptive_cw} algorithm. In \cite{q_learning_fairness}, authors developed a Q-learning solution that led to a $12\%$ fairness improvement whilst considering the delay in determining and assigning the optimal CW.

Nevertheless, these studies have shown improvements in latency, throughput, and fairness. their implementation can be quite complex as it necessitates specific changes to the current standard. Besides, many of those studies do not consider the use of HD Map data o multi-service scenarios. Furthermore, some of the required modifications may not be feasible for a multi-agent system. To address these challenges, we propose a multi-agent system where each Roadside Unit (RSU) functions as an agent, in addition to the inclusion of multi-service scenarios.

\section{Problem Statement} \label{problem_statement}
The vehicles are defined as the set $\mathcal{V} = \{1,...,N\}$, after the vehicle enters the environment it is assigned a category within the set $\mathcal{C} = \{1,...,M\}$, which denotes the type of service the vehicle will transmit.

In environments like VANET, characterized by high mobility and density, it's crucial to distribute wireless resources effectively to meet the latency and throughput needs of each service. To enhance Quality of Service (QoS), we have implemented a utility function considering both latency and throughput, as detailed in \cite{adaptive_edca}. This function provides a comprehensive evaluation of the network and ensures a balance for real-time applications that demand low latency while maintaining throughput. Moreover, it considers penalties and bonuses, as discussed in \cite{our_sojourn_single_agent}, which are applied when the latency (\ref{eq:constraint_L}) and throughput (\ref{eq:constraint_R}) constraints for each category are exceeded. Our ultimate goal is to maximize the utility function defined in eq. (\ref{eq:utility_function_penalty}). The coefficients $\alpha_1$, and $\alpha_2$ are weights to provide a trade-off between $\mathcal{R}$ and $\mathcal{L}$.

\begin{equation}
    \begin{split}
    U(c) = \alpha_1\frac{\mathcal{R}(c)}{\mathcal{R}_{max}(c)}- \alpha_2\frac{\mathcal{L}(c)}{\mathcal{L}_{max}(c)} + \mathcal{P} + \mathcal{B}
    \label{eq:utility_function_penalty}
    \end{split}
\end{equation}

Henceforth, the maximization problem for the RSU is formulated as follows:

\begin{equation}
    \underset{w_{v}}{\max}  \sum_{v \in \mathcal{V}} x_{v} U_{v}(c) \quad ,\forall c \in \mathcal{C}, \forall v \in \mathcal{V}
    \label{eq:max_problem}
\end{equation}

subject to:

\begin{equation}
    x_{v} \in \{0,1\}
    \label{eq:constraint_x}
\end{equation}
\begin{equation}
    \frac{1}{|\mathcal{V}|} \sum_{v=1}^{|\mathcal{V}|} \mathcal{L}_{v}(c) \leq \mathcal{L}_{\text{max}}(c) \quad , \mathcal{L} \in \mathbb{R}
    \label{eq:constraint_L}
\end{equation}
\begin{equation}
    \sum_{v=1}^{|\mathcal{V}|} \mathcal{R}_{v}(c) \geq \mathcal{R}_{\text{min}}(c) \quad , \mathcal{R} \in \mathbb{R}
    \label{eq:constraint_R}
\end{equation} 
\begin{equation}
    w_{v}(c) \leq w_{\text{max}}(c) \quad , w \in \mathbb{R} \quad , w \neq 0
    \label{eq:constraint_w}
\end{equation}

The constraint (\ref{eq:constraint_L}) is the maximum latency, (\ref{eq:constraint_R}) is the minimum data rate, and (\ref{eq:constraint_w}) is the maximum waiting time allowed. Finally, $x_{v}$ is the binary index that could be either 0 or 1 indicating that a vehicle is allowed to transmit.

The maximization problem at hand becomes more complex due to the involvement of multiple variables and relationships. The variables $\mathcal{L}$ and $\mathcal{R}$ exhibit direct and inverse proportional relationships with CW and the number of vehicles, respectively. Consequently, it is advantageous to employ reinforcement learning (RL) to solve the problem as it is well-suited for uncovering hidden patterns that may not easily perceptible using analytical strategies.

\section{Design} \label{design}
The paper \cite{our_sojourn_single_agent} suggests a QoS solution operating at the application layer without MAC layer modifications. In this paper, we introduced a decentralized multi-agent setup where RSUs act as agents, reducing latency by allocating waiting time for vehicles. Agents employ individual learning processes and act based on environmental feedback. The Q-learning model is compared to other updated RL algorithms such as DQN and Actor-Critic.

\subsection{Reinforcement Learning}
The RL algorithm is described in \cite{our_sojourn_single_agent}, it is a Q-learning Temporal Difference (TD), a model-free method. Thus, the agent receives the states of the environment and performs an action given by the policy $\pi(a,s)$ using $\epsilon$-$greedy$ epsilon. After the agent receives a reward $r$, the Q value table is updated using eq. (\ref{eq_q_learning}), that contains the value of the next state with a discount factor ($\gamma$), and a learning rate ($\alpha$).

\begin{equation}
    Q(s,a) = Q(s,a) + \alpha\left[r+\gamma*max_{a'}Q(s',a')-Q(s,a)\right]
\label{eq_q_learning}
\end{equation}

In \cite{our_sojourn_single_agent}, the utility function eq. (\ref{eq:utility_function_penalty}) is selected as the reward function. Thus, substituting the utility function (\ref{eq:utility_function_penalty}) in (\ref{eq_q_learning}), and for simplicity denoting $y = \gamma*max_{a'}Q(s',a')-Q(s,a)$, the proceeding eq. (\ref{eq_q_learning}) becomes, 
\vspace{0.01cm}
\begin{equation}
\begin{split}
        Q(s,a) = Q(s,a) + \alpha[U + y]
    \label{eq_q_learning_2}
\end{split}
\end{equation}

\subsubsection{DQN}
The DQN algorithm comprises several essential stages. Firstly, it requires calculating the target value using eq. \eqref{eq:dqn_target_value}. Subsequently, the loss function is computed using eq. \eqref{eq:dqn_loss_function}. Finally, the main network parameter $\theta$ is updated according to eq. \eqref{eq:dqn_network_parameter}.
\vspace{0.01cm}
\begin{equation}
\begin{split}
    y = U + \gamma*max_{a'}Q_{\theta'}(s',a')
\label{eq:dqn_target_value}
\end{split}
\end{equation} 
\vspace{0.01cm}
\begin{equation}
\begin{split}
        L(\theta) = \frac{1}{K}\sum_{i=1}^K(y_i - Q_{\theta}(s_i,a_i))^2
    \label{eq:dqn_loss_function}
\end{split}
\end{equation}
\vspace{0.01cm}
\begin{equation}
\begin{split}
        \theta = \theta - \alpha\nabla_{\theta}L(\theta)
    \label{eq:dqn_network_parameter}
\end{split}
\end{equation}

\subsubsection{Actor-Critic}
The Actor-Critic algorithm involves several key steps. Initially, the policy gradients are computed according to eq. \eqref{eq:actor_gradient}. Subsequently, the actor network parameters $\theta$ are updated based on eq. \eqref{eq:actor_network_parameter}. Following this, the loss function of the critic is computed using eq. \eqref{eq:critic_loss}. Finally, the critic network parameters $\phi$ are updated according to eq. \eqref{eq:critic_network_parameter}.

\begin{equation}
\begin{split}
       \nabla_{\theta}J(\theta) = \nabla_\theta \log\pi_\theta(a_t|s_t)(r+\gamma V_\phi(s'_t)-V_\phi(s_t))
    \label{eq:actor_gradient}
\end{split}
\end{equation}

\begin{equation}
\begin{split}
        \theta = \theta + \alpha\nabla_{\theta}J(\theta)
    \label{eq:actor_network_parameter}
\end{split}
\end{equation}

\begin{equation}
\begin{split}
       J(\theta) = (r+\gamma V_\phi(s'_t)-V_\phi(s_t))
    \label{eq:critic_loss}
\end{split}
\end{equation}

\begin{equation}
\begin{split}
        \phi = \phi - \alpha\nabla_{\phi}L(\phi)
    \label{eq:critic_network_parameter}
\end{split}
\end{equation}

\subsection{State, Action, Reward}
For the state, action, and reward, we followed the same approach as developed by us previously in \cite{our_sojourn_single_agent}. Moving on with the Q-learning design, the state, action, and reward are described in this subsection.

\subsubsection{State}
The state set $\mathcal{S}$ is defined as follows:

\begin{equation}
    S = \{Sj, Tv, C, Tcv\}
\end{equation}

$Sj$ being the sojourn time with discrete values from zero to four \cite{our_sojourn_single_agent}. It describes the time a vehicle will spend in a certain base stations. The calculation is described in \cite{our_sojourn_single_agent}. $Tv$ is the total number of vehicles active in the RSU. $C$ is the set of categories, and $Tcv$ the total number of actives vehicles per category $c$.

\paragraph{Categories}
The vehicles are designated one of the four categories randomly during the simulation, denoted by $\mathcal{C}=\{\text{VO}, \text{VI}, \text{BE}, \text{HD Map}\}$.

\subsubsection{Actions}
In this design, we have opted for eight number of actions as stated in \cite{our_sojourn_single_agent}, each corresponding to a specific waiting time before the next transmission. To map the action to each category, we have used the equation (\ref{eq:map_actions}) and Table \ref{tab:maxi_actions}. The variable $a \in \mathbb{N}$ denotes the action number, while $w_{max}$ represents the maximum waiting time specified in Table \ref{tab:maxi_actions}. Notably, the equation transforms the discrete action number into a continuous value.

\begin{equation}
    w(c) = a \cdot \left(\frac{w_{max}(c)}{|\mathcal{A}|}\right)
    \label{eq:map_actions}
\end{equation}

\begin{table}[h]
\centering
\caption{Maximum waiting time and thresholds per category \cite{our_sojourn_single_agent}}
\begin{tabular}{|c|c|c|c|}
   \hline
   Service & max wt & $\mathcal{R}$ Threshold & $\mathcal{L}$ Threshold\\
   \hline
   VO  & 0.92s & 100kpbs \cite{dataRate_cisco_voice_szigeti2005end} &  150 ms \cite{latency_cisco} \\
   \hline
   VI  & 2s & 1.25Mbps \cite{dataRateVideo_googleYouTube} & 100 ms \cite{5gaa_delay}\\
   \hline
   HD Map  & 2s & 1.25Mbps & 100 ms \cite{5gaa_delay}\\
   \hline
   BE  & 8s & 1.0Mbps  & 1000 ms \\
   \hline
   \multicolumn{2}{l}{wt = Waiting time. } \\
\end{tabular}
\label{tab:maxi_actions}
\end{table}

\subsubsection{Reward}
The utility function (\ref{eq:utility_function_penalty}) will be utilized to calculate the reward, while taking into account the constraints for each of the categories and their respective penalties and bonuses. The constraints are linked with to the threshold values registered on Tables \ref{tab:maxi_actions}. The utility, the values of 0.3 and 0.7 for $\alpha_1$, and $\alpha_2$ respectively, have been selected after exhaustive simulation. The allocation of these values revealed that the data rate is maintained while obtaining lower latency.

\section{Simulation} \label{simulation}
To evaluate the solution, OMNet++ version 6 \cite{omnetppOMNeTDiscrete} is utilized as the network simulator, plus the library Inet frameworks \cite{omnetppINETFramework}. The open source simulator and library provide the Physical, and MAC layer stack to simulate a wireless network using IEEE802.1p. Regarding the traffic flow, we use veins that connect OMNet++ with SUMO a continuous traffic simulation. Finally, for the machine learning, we opted to utilize and accommodate the solution developed by M. Schettler et al \cite{veinsgym} to connect our RL algorithm to the OMNet++ simulation environment. Please refer to Table \ref{table:simulation_parameters} our simulation parameters.

The traffic flow generation is calculated by the number of vehicles on a heavy traffic road and rush hour. This information is extracted from the Traffic and Accident Data Unit / North East Regional Road Safety Resource \cite{dataset}. Thus by dividing the total number of vehicles per hour, the entry interval is $0.66$s.

During the simulation, RSU functions as an autonomous agent, exercising independent decision-making capabilities. Information exchange occurs solely through the observation of the environment state, facilitating a decentralized decision-making process. 

The analysis of the simulation is performed in terms of latency, and throughput Key Performance Indicators (KPIs). Latency is computed by subtracting the difference between the generated packet time, and the arrival time at the destination. Fairness is introduced only in the comparison between RL approaches. The formula to express fairness is Jain's fairness index \cite{fairness_formula_jain}.

\begin{table}[ht]
\caption{Simulation Parameters \cite{our_sojourn_single_agent}}
\begin{center}
\begin{tabular}{|c|c|}
\hline
\textbf{\textit{Parameter}} & \textbf{\textit{Value}} \\
\hline
Tile Dimension & 300x100 meters  \\
\hline
Episodes & 50  \\
\hline
$\epsilon-greedy$ & 0.2  \\
\hline
Discount Factor $\gamma $& 0.99  \\
\hline
Learning Rate $\epsilon $& 0.1  \\
\hline
$\alpha_1 $, $\alpha_2 $ & 0.3 and 0.7 respectively  \\
\hline
Simulation time & 250s  \\
\hline
Vehicle Density$^a$ & varies according traffic flow  \\
\hline
Coverage Area & 200m  \\
\hline
Vehicle max speed & 17m/s  \\
\hline
Vehicle Acceleration  & {2.6 m}/{$s^2$} \cite{acceleration_deceleration_petrol_car} \\
\hline
Vehicle Deceleration  & {4.5 m}/{$s^2$} \cite{acceleration_deceleration_petrol_car}  \\
\hline
Tx Power & 200 mW  \\
\hline
Frequency & 5.9 GHz  \\
\hline
Bandwidth & 10MHz  \\
\hline
Best-effort data rate & 28Mbps  \\
\hline
HD Map data rate & 4Mbps \cite{5gaa_delay}\\
\hline
Video data rate & 5Mbps \cite{dataRateVideo_googleYouTube} \\
\hline
Voice data rate & 100kbps \cite{dataRate_cisco_voice_szigeti2005end} \\
\hline
TXOP limit & Disabled as per standard  \\
\hline
Number of hidden layers & 1 \\
\hline
Number of neurons in hidden layer & 32 \\
\hline
Buffer size & 500  \\
\hline
\end{tabular}
\label{table:simulation_parameters}
\end{center}
\end{table}

In our simulation, while comparing DQN and Actor-Critic to Q-learning, we ensured a fair comparison by using identical learning rates and discount factors across all algorithms. Moreover, both DQN and Actor-Critic were configured with same number of hidden layers in their neural networks. This deliberate choice offers valuable insights into the level of optimization required by DQN and Actor-Critic compared to the more straightforward Q-learning algorithm in the context of VANETs.

\section{Result and Discussion} \label{results}
The results of the suggested Q-Learning TD RL are contrasted with the existing standard IEEE802.11p in various scenarios, including standard without QoS, and with QoS, new AC HD Map, and the single agent model without the incorporation of penalties and bonuses.

\subsection{Validation of the model} 
We conducted various tests to evaluate the model's robustness across diverse scenarios. Initially, we configured each service with distinct packet sizes. Then, we investigated advanced RL algorithms such as deep Q-Network (DQN) and Actor-Critic. Finally, we subjected the model to tests in a multi-RSU setting, with traffic generation extracted from the dataset \cite{dataset} to measure its performance.

\subsubsection{Adjusting different Packet Sizes for each Category}

When considering different packet sizes, we ensured a consistent sending interval for each category: $10$ms for voice, $1$ms for video, $1.2$ms for HD map, and $250$us for best-effort data. However, the packet size has been changed from $900$Bytes to $125$Bytes for voice, $600$Bytes for video, $600$Bytes for HD map, and $900$Bytes for Best-effort data. These adjustments were made while preserving the data rate requirements.

\begin{figure}[h]
\centering
\includegraphics[width=\linewidth]{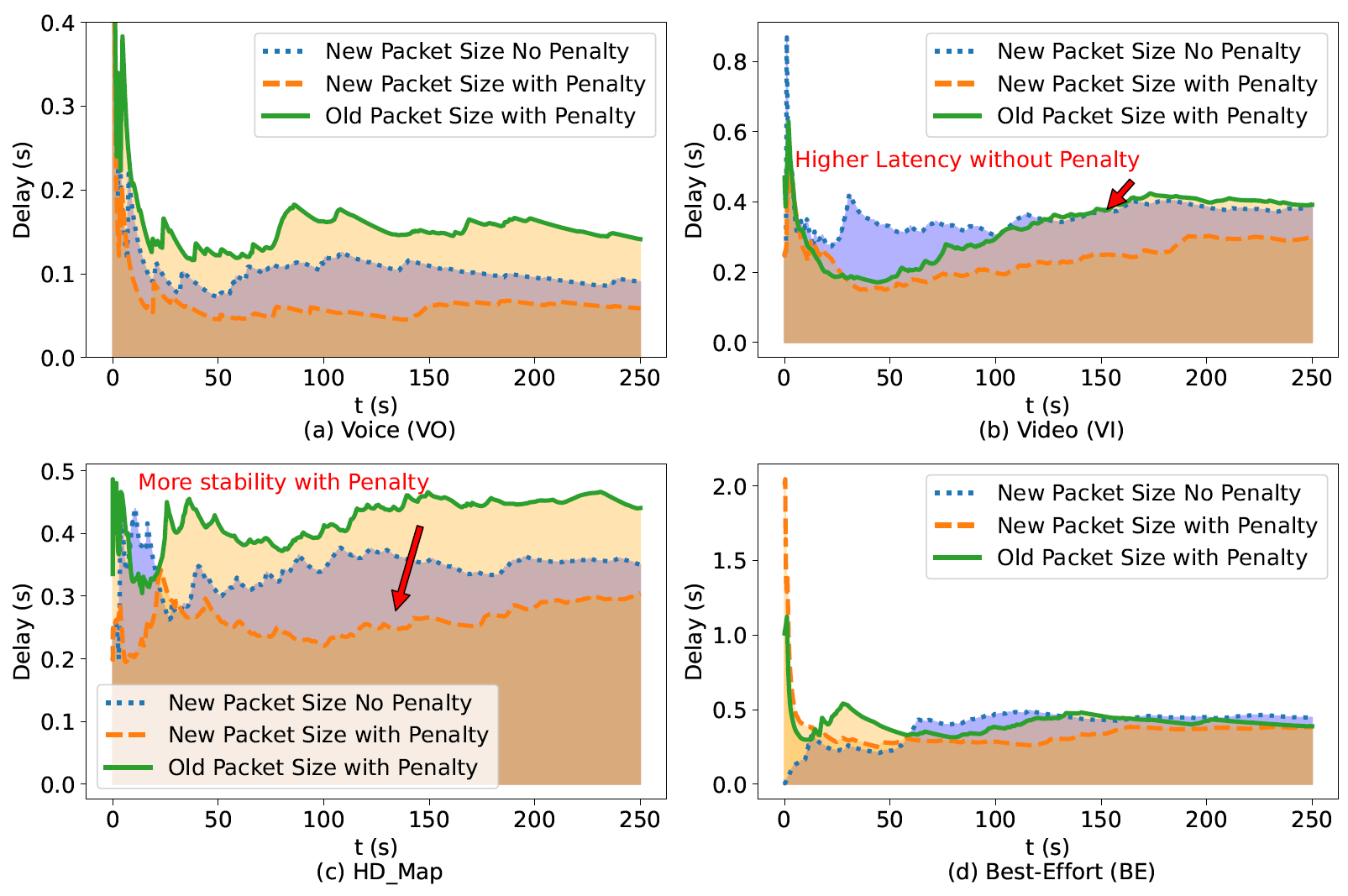}%
\caption{Latency comparison between solution with and without penalties using different packet size. (a) HD Map, (b) Video, (c) Best-Effort, and (d) Voice.}
\label{fig:diff_packet_size_latency}
\end{figure}

\begin{figure}[h]
\centering
\includegraphics[width=\linewidth]{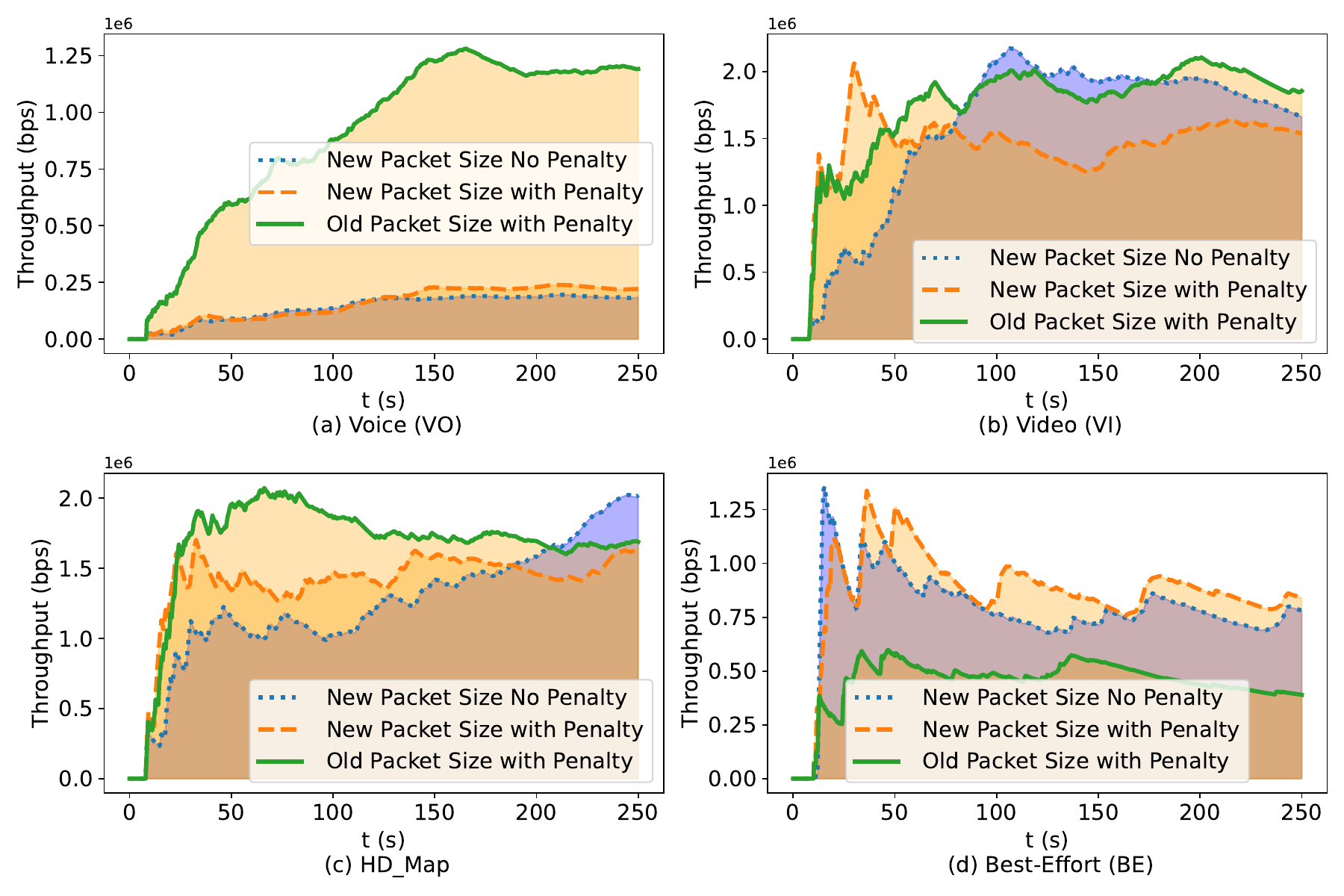}%
\caption{Throughput comparison between solution with and without penalties using different packet size. (a) HD Map, (b) Video, (c) Best-Effort, and (d) Voice.}
\label{fig:diff_packet_size_throughput}
\end{figure}

From Fig. \ref{fig:diff_packet_size_latency}, and \ref{fig:diff_packet_size_throughput}, as expected the average delay and throughput were reduced due to a lower packet size. Results are more stable and align with the threshold while considering penalties which enhanced the performance by approximately $50\%$ in terms of delay for voice Fig. (\ref{fig:diff_packet_size_latency} (a)), video Fig. (\ref{fig:diff_packet_size_latency} (b)), and HD Map (Fig. \ref{fig:diff_packet_size_latency} (c)) while maintaining the average throughput constant throughout the simulation as seen in \ref{fig:diff_packet_size_throughput}.

\subsubsection{Comparison RL Complexity Algorithm}
\begin{figure}[h]
\centering
\includegraphics[width=\linewidth]{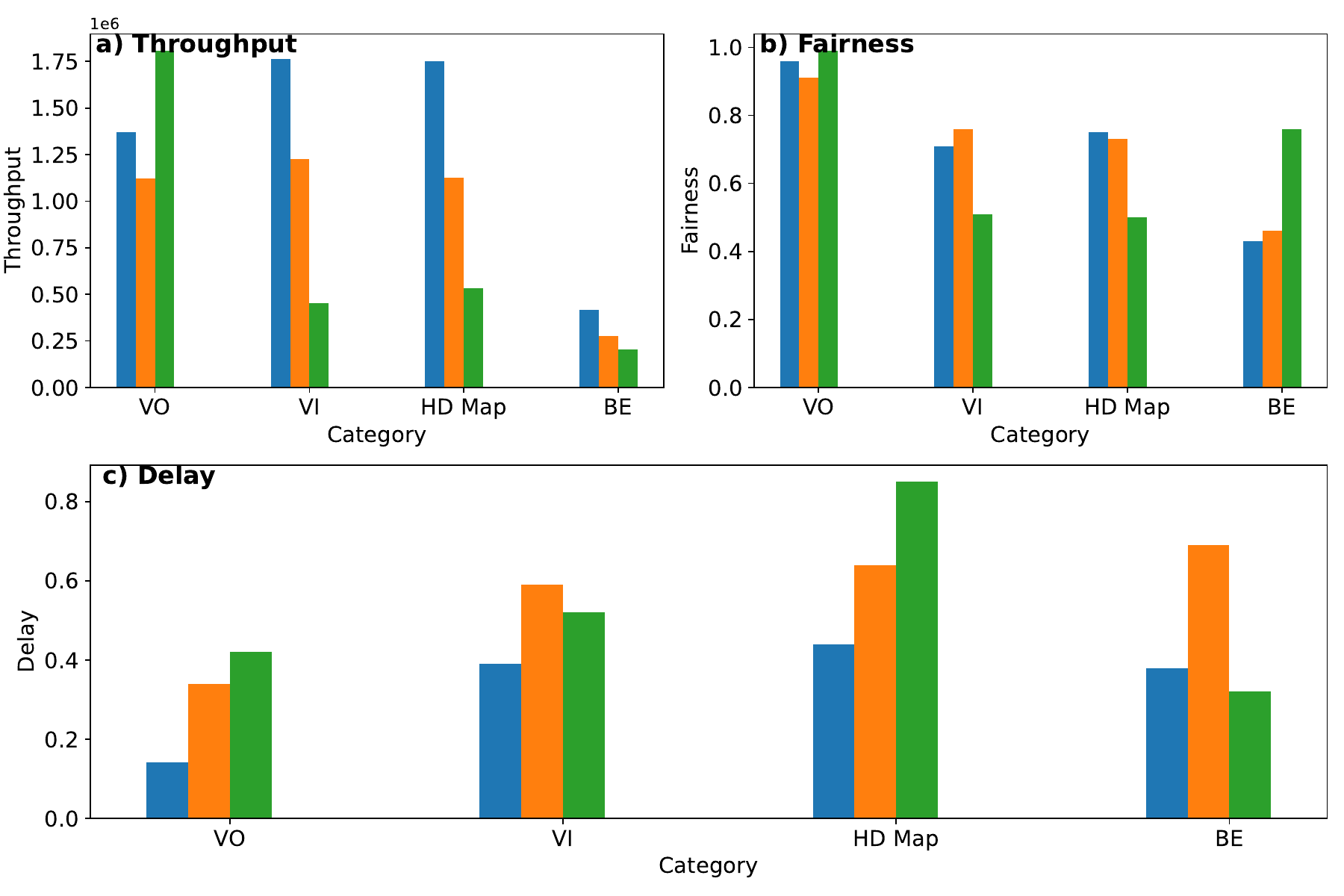}%
\caption{Comparison with different reinforcement learning algorithms using IEEE802.11p}
\label{fig:diff_RL_algorithms}
\end{figure}

In addition, we conducted a comparison of various RL methods, including DQN and Actor-Critic, we kept all states, hyper-parameters, and actions the same as the Q-learning design. The results presented in Fig. \ref{fig:diff_RL_algorithms} reveal that both DQN and Actor-Critic require extra fine-tuning work to achieve comparable or better results than the Q-learning algorithm proposed in \cite{our_sojourn_single_agent}. For instance, DQN results exhibited similar behavior to Q-learning in terms of the level of priorities; However, it experienced a delay increase of $133\%$ in voice, $50\%$ in video and HD map, and $71\%$ in best-effort. Despite this, the DQN approach demonstrated a throughput performance closer to the threshold of $1.25$Mbps.

Similarly, the Actor-critic algorithm exhibited comparable behavior but yielded lower performance than the DQN algorithm. If we analyze fairness the Actor-critic provided the lowest fairness for VI and HD Map, and the highest for BE which is not desirable. Overall, it is evident that implementing a more complex RL algorithms requires a thorough examination in tuning hyper-parameters, exploration-exploitation, and state selection to attain a better performance compared to our Q-learning solution. This demonstrated, that our solution which requires lower computational capacity offers a better performance.

\subsubsection{Multi RSU in Real Environment}
\begin{figure}[h]
\centering
\includegraphics[width=2in,height=2in]{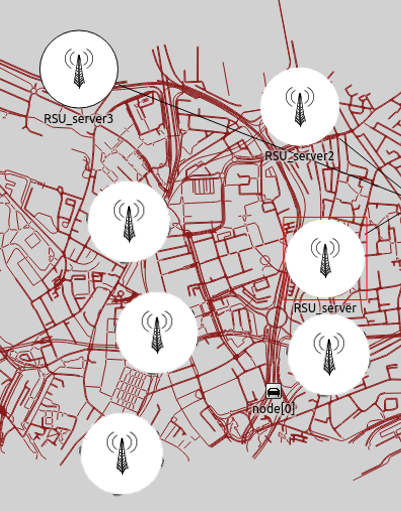}%
\caption{Scenario Multi-RSU.}
\label{fig:scenario_multi_rsu}
\end{figure}

Finally, we opted to evaluate the model in a more realistic environment, selecting the city of Newcastle within the UK as our city of choice. Our traffic data was supplied by Traffic and Accident Data Unit/North East Regional Road Safety Resource \cite{dataset}. As shown in Fig. \ref{fig:scenario_multi_rsu}, we had a set of 7 RSUs configured with the technology IEEE802.11p. The AVs select the RSU based on the shortest distance. Firstly, the simulation was performed using the standard with and without EDCA. Secondly, the new AC for HD map solution was also evaluated. Afterward, the previously trained model was evaluated. Finally, the model was trained for ten episodes in two scenarios. The first scenario used a single agent, while the second used a multi-agent system with each RSU acting as an agent.

\begin{figure}[h]
\centering
\includegraphics[width=\linewidth]{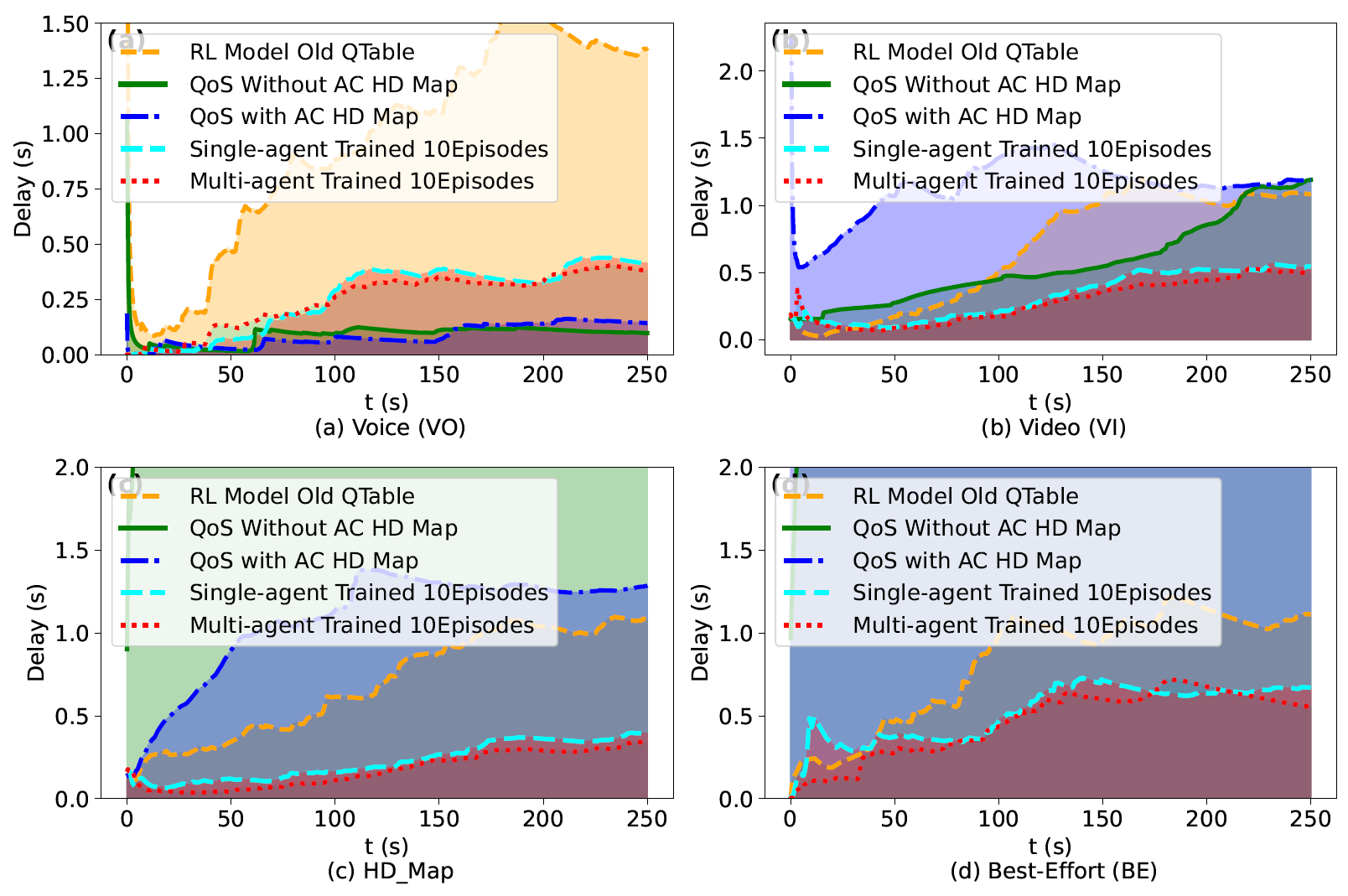}%
\caption{Latency for MultiRSU environment. (a) Voice, (b) Video, (c) HD Map, and (d) Best-Effort.}
\label{fig:comparison_latency_newcastle}
\end{figure}

\begin{figure}[h]
\centering
\includegraphics[width=\linewidth]{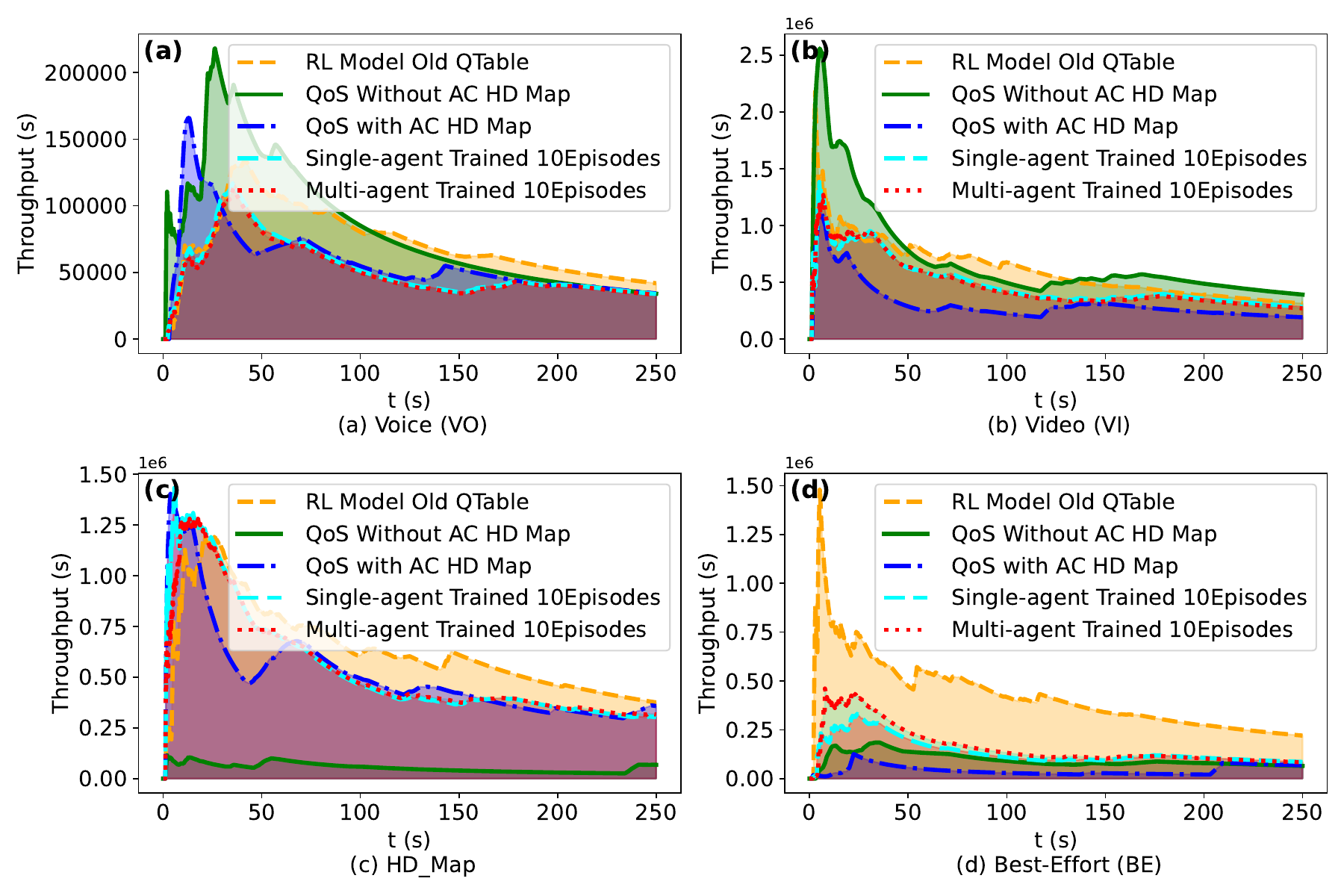}%
\caption{Throughout MultiRSU environment. (a) Voice, (b) Video, (c) HD Map, and (d) Best-Effort.}
\label{fig:comparison_throughput_newcastle}
\end{figure}

As shown in Fig. \ref{fig:comparison_latency_newcastle}, and \ref{fig:comparison_throughput_newcastle}, the single-agent and multi-agent trained for ten episodes offer lower latency compared to the currently trained model and the standard for Video, HD map, and best-effort. Regarding the delay, the voice service with the new training solution exceeded the old trained model by $50\%$. In addition, the multi-agent (trained by ten episodes) improves the delay by $10\%$ compared with the single agent (trained by ten episodes). For video and HD Map, there is an improvement of approximately $70\%$ against the old trained model. However, for the voice service, the standard EDCA presents a lower delay compared to the multi-agent model with a total average delay of $0.1$s and $0.32$s, respectively.

Overall, the previous results regarding multi-RSU provides insights to start our future work with the implementation of a more complex RL algorithm and multi-agent system to mitigate the deterioration of the network performance.

\section{Conclusion} \label{conclusion}
Our research has resulted in the development of a multi-agent system for a multi-RSU scenario. It has shown that the single agent solution in \cite{our_sojourn_single_agent} is transferable and could be re-trained to improve the network performance in terms of latency and throughput. Additionally, it reveals that the standard IEEE802.11p is insufficient to support a real-time HD map system in a real-world environment without adapting the packet size of each service. Nevertheless, it demonstrated that the reward function selected provides enough feedback in a multi-agent scenario to learn an optimal policy without communication between the agents directly. This is favorable as we avoid the misuse of the wireless channel. Also, the Q-learning solution demonstrated more efficiency. The latency and throughput were improved compared with a more complex and computationally-intensity RL algorithm such as DQN, and Actor-critic. It is important to recall, that there was no optimization performed on DQN and Actor-Critic algorithm. This outcome emphasizes the solution's potential for achieving desirable results with comparatively less tuning and optimization effort. It also indicates its viability for real-world deployment. Nevertheless, future endeavors may require adopting a multi-agent approach with more sophisticated reinforcement learning algorithms.

\section*{Acknowledgment}
This work was partly funded by EPSRC with RC Grant reference EP/Y028023/1, UKRI under grant number EP/Y028023/1, the European Horizon2020 MSCA programme under grant agreement  No. 101086228.

\ifCLASSOPTIONcaptionsoff
  \newpage
\fi

\bibliographystyle{IEEEtran}
\bibliography{IEEEabrv,Bibliography}
\end{document}